\documentclass{article}
\usepackage{booktabs}
\usepackage[T1]{fontenc}
\usepackage[utf8]{inputenc}
\usepackage{ismir} % Remove the "submission" option for camera-ready version
\usepackage{amsmath,cite,url}
\usepackage{multirow}
\usepackage{booktabs}
\usepackage{tabularx}
\usepackage{graphicx}
\usepackage{caption}
\usepackage{color}

\usepackage{float}

\title{MIDI-VALLE: Improving Expressive Piano Performance Synthesis Through Neural Codec Language Modelling}

\multauthor
  {Jingjing Tang$^1$\hspace{1cm} Xin Wang$^2$\hspace{1cm} Zhe Zhang$^2$\hspace{1cm}}
  {{\bf Junichi Yamagishi$^2$\hspace{1cm} Geraint Wiggins$^{1,3}$\hspace{1cm} Gy\"orgy Fazekas$^1$}\\
  $^1$Centre for Digital Music, Queen Mary  University of London, UK\\
  $^2$National Institute of Informatics, Japan\\
  $^3$Vrije Universiteit Brussel, Belgium\\
  {\tt\small jingjing.tang@qmul.ac.uk}
  }

\def\authorname{J. Tang, X. Wang, Z. Zhang, J. Yamagishi, G. Wiggins and G. Fazekas}

\begin{document}

\maketitle
\begin{abstract}
Generating expressive audio performances from music scores requires models to capture both instrument acoustics and human interpretation. Traditional music performance synthesis pipelines follow a two-stage approach, first generating expressive performance MIDI from a score, then synthesising the MIDI into audio. However, the synthesis models often struggle to generalise across diverse MIDI sources, musical styles, and recording environments. To address these challenges, we propose MIDI-VALLE, a neural codec language model adapted from the VALLE framework, which was originally designed for zero-shot personalised text-to-speech (TTS) synthesis. For performance MIDI-to-audio synthesis, we improve the architecture to condition on a reference audio performance and its corresponding MIDI. Unlike previous TTS-based systems that rely on piano rolls, MIDI-VALLE encodes both MIDI and audio as discrete tokens, facilitating a more consistent and robust modelling of piano performances. Furthermore, the model’s generalisation ability is enhanced by training on an extensive and diverse piano performance dataset. Evaluation results show that MIDI-VALLE significantly outperforms a state-of-the-art baseline, achieving over 75\% lower Fréchet Audio Distance on the ATEPP and Maestro datasets. In the listening test, MIDI-VALLE received 202 votes compared to 58 for the baseline, demonstrating improved synthesis quality and generalisation across diverse performance MIDI inputs.

%As demonstrated by the evaluation results, our model not only inherits zero-shot capabilities but also effectively adapts to a wide range of performance MIDI inputs while faithfully preserving expressive timing, dynamics, and the acoustic qualities of the reference performance.
\end{abstract}

\section{Introduction}\label{sec:introduction}
Music performance synthesis (MPS) refers to the process of generating expressive audio performances from music scores. This task requires models to capture acoustic characteristics of musical instruments and infuse human-like expressiveness into music scores. While considerable progress has been made in modelling these aspects separately, an effective MPS system is expected to integrate both dimensions to achieve high-quality synthesis.

A common approach to MPS involves a two-stage pipeline consisting of an expressive performance rendering (EPR) model, which generates expressive performance MIDI from a score, and an expressive performance synthesis (EPS) model, which converts performance MIDI into audio \cite{wu2021midi, dong2022deep, tang2025towards}. In recent works for developing EPS models \cite{cooper2021text, shi2023can, dong2022deep, tang2025towards}, the task has been recognised as analogous to speech synthesis, as both generate audio from symbolic representations. This parallel motivated researchers to apply advanced techniques from the text-to-speech (TTS) domain to address the challenges in EPS. Previous studies have demonstrated the effectiveness of TTS techniques, such as WaveNet \cite{hawthorne2018enabling} and acoustical models with vocoders \cite{cooper2021text, shi2023can, dong2022deep}, in synthesising performance MIDI to audio. However, due to limited training data diversity and constrained architecture design, these models struggle to generalise across acoustic environments and timbre variations, limiting the expressiveness and realism of their outputs. Moreover, when integrating these EPS systems with EPR models, discrepancies in the way EPR and EPS models process and represent MIDI data introduce inconsistencies. These discrepancies often result in the loss of fine-grained temporal details, leading to the reduced synthesis quality.

To address the limitations, we introduce a novel EPS model, MIDI-VALLE\footnote{Demo and codes are available at \url{https://tangjjbetsy.github.io/MIDI-VALLE/}}, adapted from VALLE, a state-of-the-art TTS framework for zero-shot personalised speech synthesis \cite{chen2025valle}. The VALLE model conditions synthesis on speaker-specific audio prompts, enabling zero-shot adaptation to unseen speakers. We optimise this architecture for performance MIDI-to-audio synthesis by conditioning on a reference audio performance and its corresponding MIDI representation. Instead of using the Maestro dataset \cite{hawthorne2018enabling}, which contains recorded performance MIDI and audio pairs, we train the MIDI-VALLE on ATEPP \cite{zhang2022atepp}, a larger and more diverse dataset comprising transcribed performance MIDI and audio pairs. This allows the model to learn from a broader range of musical expressions, improving generalisation across unseen MIDI sources, composition styles, and recording environments. As demonstrated by both objective and subjective evaluation results, MIDI-VALLE shows enhanced adaptability and robustness in handling diverse performance inputs compared to previous state-of-the-art EPS models,

\begin{figure*}[!hbt]
    \centering
\includegraphics[width=.96\textwidth]{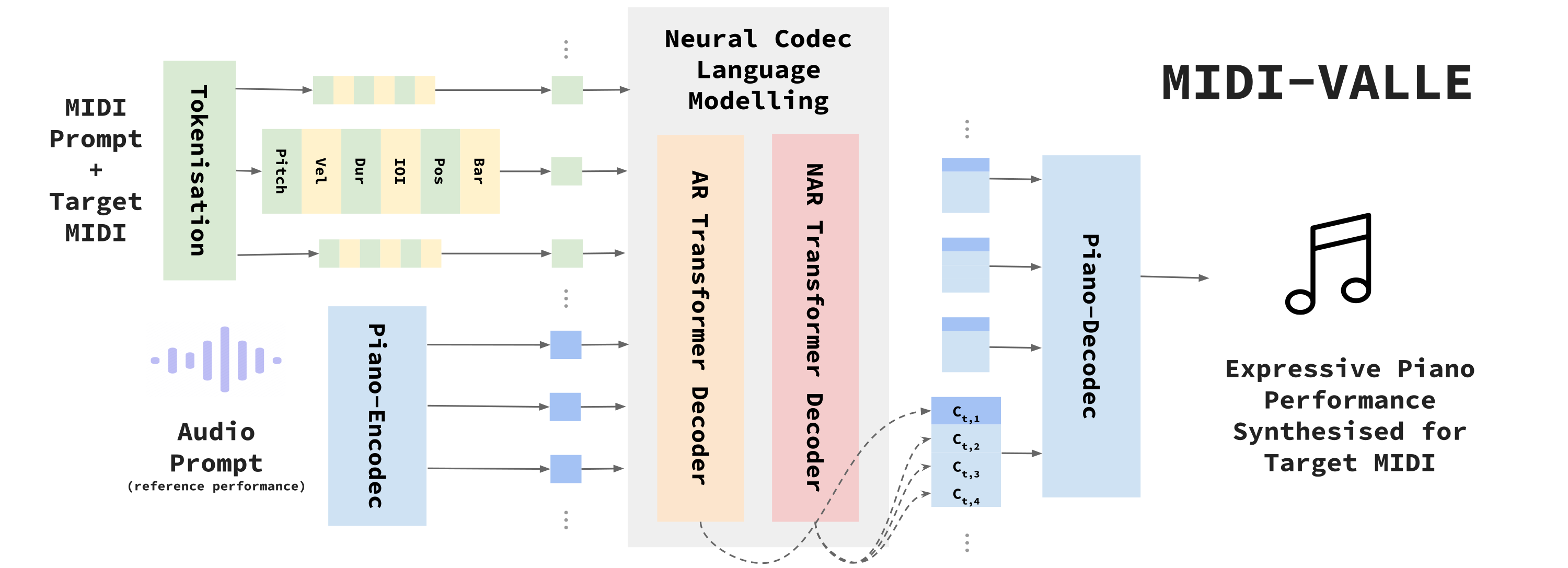}
    \caption{Overview of the MIDI-VALLE architecture, adapted from VALLE \cite{chen2025valle}. Audio prompt is a 3-second segment selected from a reference performance. The text prompt in VALLE is replaced with the corresponding MIDI prompt concatenated with the target MIDI for synthesis.}
    \label{fig:architecture}
    \vspace{-1.5em}
\end{figure*}

Moreover, to fully leverage neural codec language modelling, we tokenise performance MIDI and audio using Octuple MIDI tokenisation method \cite{zeng-etal-2021-musicbert} and a high-fidelity audio codec model \cite{fossez2023high}, which ensures accurate reconstruction from audio tokens. Compared to traditional piano-roll and spectrogram representations, this discrete token-based approach ensures a more consistent alignment between MIDI and audio. The results from the listening test demonstrate that MIDI-VALLE, when integrated with different EPR models in a two-stage MPS pipeline, provides a more robust and adaptable synthesis framework.

\section{Related Works}
\subsection{Expressive Performance Synthesis}
\label{sec:rw_eps}
In the EPS domain, several studies have explored various approaches for MPS, including DDSP-based modelling \cite{renault2023ddsp, wu2021midi} and TTS-inspired models \cite{cooper2021text, shi2023can, dong2022deep, tang2025towards}. These TTS-inspired models typically process piano performance MIDIs as piano rolls for audio synthesis. Hawthorne et al.~\cite{hawthorne2018enabling} employed WaveNet to map piano rolls directly to waveforms. More recent works \cite{cooper2021text, shi2023can, dong2022deep, tang2025towards} adapted transformer-based TTS models \cite{li2019neural, fast2019ren} to first convert piano rolls into intermediate acoustic representations, such as spectrograms. These representations were subsequently transformed into waveforms using vocoders like HiFi-GAN \cite{kong2020hifi}. These EPS models were mainly trained on the Maestro dataset \cite{hawthorne2018enabling}, which consists of recorded MIDI and audio pairs from piano competitions. Although the dataset includes performances of diverse compositions, they were recorded in a relatively homogeneous acoustic environment. This lack of acoustic variety limits the ability of models trained on the dataset to generalise to more varied acoustic conditions. Tang et al.~\cite{tang2025towards} attempted to fine-tune a state-of-the-art model \cite{shi2023can} using the ATEPP \cite{zhang2022atepp} dataset, which features recordings captured in a broader range of acoustic settings. However, the fine-tuned model still struggled to produce consistent ambient sounds, applying mismatched or inconsistent room reverberation and background noise. 
% It also had difficulty accurately rendering the natural timbre and spatial characteristics of the piano during synthesis.

A key challenge in creating a two-stage pipeline for MPS is the difference in MIDI representations used by EPR and EPS models, particularly in temporal information. EPS models typically use piano-roll representations, while EPR models either tokenise MIDI \cite{tang2023reconstructing, tang2025towards, borovik2023scoreperformer} into discrete events or encode continuous features \cite{renault2023unpair, Jeong2019VirtuosoNetAH, zhang2024dexter} like timing and velocity. These differences complicate MIDI conversion between stages, especially with note timing and pedal treatment. Consequently, performance MIDI generated from EPR models differs significantly from the performance MIDI used in EPS models, making direct integration impractical without additional fine-tuning. More details are discussed in Section~\ref{sec:listening_test} and Section~\ref{sec:results}.

% Some models omit pedal data \cite{renault2023unpair, borovik2023scoreperformer, tang2025towards}, while others predict pedal positions or adjust note durations for pedal-off events \cite{Jeong2019VirtuosoNetAH, zhang2024dexter, cooper2021text, shi2023can}. These inconsistencies 

\subsection {Neural Codec Language Modelling for Audio Generation}
Recent advances in audio and music generation have leveraged neural codec language models to address the challenges of generalising across diverse acoustics and music styles. The state-of-art text-to-audio \cite{borsos2023audiolm} and text-to-music \cite{copet2023simple, agostinelli2023musiclm} models use codec models like Encodec \cite{fossez2023high} and SoundStream \cite{zeghidour2021soundstream} to compress audio into discrete tokens, enabling more efficient training on large-scale datasets by reducing computational costs. In the TTS domain, the VALLE model \cite{chen2025valle}, inspired by AudioLM \cite{borsos2023audiolm}, uses Encodec to synthesise high-quality speech while preserving speaker-specific features. By replacing mel-spectrograms with compressed audio codec tokens, VALLE formulates TTS as conditional codec language modelling. This enables effective zero-shot timbre adaptation and preservation of speaker emotion and the acoustic environment encoded in the reference prompt. Building on this approach, we extend codec language modelling to piano performance synthesis by tokenising both performance MIDI and audio, demonstrating the effectiveness of codec language modelling in synthesising expressive piano performances.

\section{MIDI-VALLE for Piano Synthesis}
\label{sec:model}
Our MIDI-VALLE model focuses on performance MIDI-to-audio synthesis, drawing parallels to text-to-speech synthesis by VALLE. The following sections discuss the tokenisation strategies and key architectural differences between MIDI-VALLE and VALLE, highlighting the similarities and distinctions between speech and music synthesis.

\subsection{Tokenisation}
\subsubsection{Audio Tokenisation}
Instead of the original Encodec model \cite{fossez2023high}, we follow the audio tokenisation approach applied in MusicGen \cite{copet2023simple}. Specifically, we fine-tune a four-level residual vector quantisation (RVQ) \cite{gray1984vq} to generate four codebooks that represent the audio samples. In RVQ, each quantiser encodes the residual error from the previous one, creating interdependencies among the codebooks. As observed in \cite{valle, fossez2023high},  the first codebook encodes the primary acoustic information, while the subsequent codebooks refine the output by modelling finer details. The fine-tuned codec, \textbf{Piano-Encodec}, converts audio performances into discrete tokens while preserving high-fidelity acoustics and timbral characteristics. The decoder then reconstructs the audio from these tokens.

\subsubsection{MIDI Tokenisation}
\label{sec:midi_tokenisation}
Classical piano music and speech differ significantly in complexity and structure. Classical music features intricate frequency patterns and precise timing, making segmentation challenging due to issues with note separation, timing accuracy, and managing prolonged note durations caused by pedalling. In contrast, speech is simpler, with clear segmentation based on phoneme boundaries and greater tolerance for timing variations. 

\begin{table}[htb]
\centering
\begin{tabular}{cccccc|c}
\hline
\multicolumn{6}{c|}{\textbf{Music}} & \textbf{Speech} \\ 
\hline
Pitch & Vel & Dur & IOI & Pos & Bar & \multirow{2}{*}{} \\
\cline{1-6}
92 & 68 & 1156 & 772 & 388 & 20 & \multirow{-2}{*}{512} \\
\hline
\end{tabular}
\caption{Vocabulary sizes for musical features and speech.}
\label{tab:token_numbers}
\end{table}

We employ the Octuple MIDI tokenisation method \cite{zeng-etal-2021-musicbert}, as utilised in \cite{tang2025towards}, to achieve a consistent and discrete representation of piano performances within the EPR and EPS systems. Unlike methods such as Compound Word \cite{hsiao2021compound} or REMI \cite{huang2020pop}, the Octuple approach uses distinct vocabularies for each musical feature, enabling note-wise encoding and resulting in a $K \times N$ array (number of features $\times$ number of notes). This method reduces vocabulary size and structural complexity, resulting in shorter token sequences without needing to group note features \cite{zeng-etal-2021-musicbert}. We extend the Octuple method by tokenising the inter-onset interval (IOI) to capture onset timing differences between consecutive notes. The MIDI tokenisation offers advantages over the piano-roll representation used in prior studies \cite{renault2023ddsp, shi2023can, cooper2021text, tang2025towards}. Piano-roll encodes only note onsets and durations on a fixed temporal grid, lacking the resolution and flexibility to capture subtle timing variations that significantly influence articulation.

Table~\ref{tab:token_numbers} illustrates the structural and representational differences between MIDI and text tokens. MIDI tokens comprise multiple sequences that encode musical features such as pitch, velocity (Vel), duration (Dur), inter-onset interval (IOI), position (Pos), and bar, explicitly representing timing information. These sequences are processed through different embedding layers and concatenated for embedding pooling \cite{hsiao2021compound}. In contrast, speech text is tokenised by a single sequence of integers representing phonemes, with timing implicitly conveyed through token order. These differences in token representation are critical to the successful training of MIDI-VALLE.

\subsection{Model Design}
Unlike VALLE, MIDI-VALLE is designed to preserve the timbral and acoustic characteristics of the reference piano performance. Given a piano performance dataset $D_p = \{x_i, y_i\}$, where $x = \{x_0, x_1, ..., x_L\}$ is a MIDI token sequence and $y$ is the corresponding audio segment, the audio is encoded into discrete acoustic codes using the pre-trained Piano-Encodec model: $encodec(y) = C_{T \times 4}$, where $C$ is a two-dimensional codec matrix representation and $T$ is the codec sequence length. During training, the model learns to predict a codec matrix $\hat{C}$ from the input MIDI $x$, and an acoustic prompt matrix $\tilde{C}$ which is derived from the first three seconds of the corresponding audio. The synthesised audio is reconstructed by $\hat{y} = \text{decodec}(\hat{C})$, aiming to approximate the original audio $y$. The model is trained to maximise the conditional likelihood $\max p(C \mid x, \tilde{C})$.

As shown in Figure~\ref{fig:architecture}, MIDI-VALLE follows a similar design in VALLE, comprising an autoregressive (AR) transformer decoder that predicts discrete tokens from the first quantiser, $ c_{t,1} $, and a non-autoregressive (NAR) transformer decoder that generates codes for the remaining three quantisers, $ c_{t,2:4} $. To process MIDI token sequences, both models employ the embedding pooling \cite{hsiao2021compound} technique to map concatenated embeddings of various musical features to match the required input size. The AR decoder takes the MIDI token sequence as input and autoregressively predicts audio codec tokens in a causal manner, without using any acoustic prompts. In contrast, the NAR decoder is conditioned on an acoustic prompt 
$\tilde{C}$, extracted from the first three seconds of the performance. This replaces the neighbouring-context strategy used in VALLE and better preserves musical coherence, as acoustic characteristics can change rapidly and vary significantly between segments. During NAR training, each token in the self-attention layer can attend to all input tokens. Despite their different decoding approaches, both the AR and NAR models share the same architecture: 12 attention layers, 16 attention heads, and hidden dimensions of size 1024.

During inference, the model inputs a target MIDI for synthesis and optionally accepts an audio prompt with its corresponding MIDI segment as the MIDI prompt. The audio prompt could be any 3-second excerpt from any recorded performance, and the associated MIDI prompt is concatenated to the beginning of the target MIDI before tokenisation. The impact of selecting different prompts is discussed in Section~\ref{sec:results_sbj}. The encoded audio prompt, if provided, is appended after the MIDI tokens in the input to the AR decoder for controlling the acoustic environment and timbral characteristics. The model then estimates the audio codec tokens for the target MIDI and reconstructs the corresponding audio performance using the Piano-Encodec. %This process synthesises a performance that aligns with the musical content of the target MIDI while preserving the acoustic environment and timbral characteristics of the audio prompt.
\vspace{-.4em}

\section{Experiments}
\subsection{Datasets}
We used the ATEPP \cite{zhang2022atepp} dataset, excluding low-quality performances and their transcriptions. A total of 8,825 performance recordings were selected and split into training, validation, and test sets in an 8:1:1 ratio. The repertoire has around 700 hours of audio recordings from 1,099 albums, featuring 1,523 compositions by 25 composers, performed by 46 pianists. All the performances were segmented randomly into clips of 15-20 seconds. To ensure precise alignment between the audio segments and the corresponding MIDIs, notes were truncated at segmenting points, with the remainder continuing in the next segment if a note was interrupted. Due to limited pedalling transcription accuracy in the ATEPP dataset, pedal information was excluded during MIDI tokenisation, and note durations represent raw durations only, without sustain extension.

\subsection{Implementation Details}
The Encodec model \cite{copet2023simple} was fine-tuned using audio from the ATEPP dataset. All performances were converted into 32kHz monophonic audio and encoded with a frame rate of 50 Hz. The extracted audio embeddings were quantised using RVQ with four quantisers, each having a codebook size of 2048. One-second audio segments were randomly sampled from the entire ATEPP dataset at each epoch, following the strategy proposed in \cite{fossez2023high}. Fine-tuning was carried out over 40 epochs on a Tesla A100 GPU for one day, with performance improvements discussed in Section~\ref{sec:results_obj}.

Our MIDI-VALLE was implemented based on an unofficial version of VALLE \cite{valle}, with training optimised using the ScaledAdam \cite{yao2024zipformer} optimiser and a base learning rate of 0.05. The learning rate was adjusted using the Eden scheduler, as described in \cite{yao2024zipformer}. The AR and NAR decoders were trained jointly, with gradients updated in the same step, converging after approximately 300k steps (2.5 days) on two Tesla A100 GPUs.
\section{Evaluation}
\subsection{Objective Metrics}
To evaluate the performance of the proposed MIDI-VALLE system, we employ three objective metrics: Fréchet Audio Distance (FAD) \cite{Kilgour2019FrchetAD, fadtk}, spectrogram distortion, and chroma distortion. FAD measures the perceptual quality and realism of generated audio by comparing it to reference performances using embeddings extracted from Piano-Encodec. Adapted from \cite{tang2025towards}, spectrogram distortion evaluates the fidelity of reconstructed acoustics and timbre, while chroma distortion evaluates harmonic consistency at the pitch class level. They are computed using the normalised root mean square error and mean absolute error, respectively.

\begin{table}[htb]
\centering\small
\begin{tabular}{c|c|c|c}
\hline
\textbf{Dataset} & \textbf{Genre} & \textbf{MIDI Type} & \textbf{RE$^\star$} \\
\hline
ATEPP \cite{zhang2022atepp} & classical & Transcribed & Live \& Studio \\
Maestro \cite{hawthorne2018enabling} & classical & Recorded & Competition \\
Pijama \cite{edwards2023pijama} & jazz & Transcribed& Live \& Studio \\
\hline
\end{tabular}
\caption{Comparison of the three piano solo datasets used for evaluation. $^\star$RE stands for recording environment.}
\label{tab:dataset_comparison}
\vspace{-.4em}

\end{table}

We evaluate MIDI-VALLE against the state-of-the-art TTS-based EPS system, M2A \cite{tang2025towards}, using three datasets: ATEPP \cite{zhang2022atepp}, Maestro \cite{hawthorne2018enabling}, and Pijama \cite{edwards2023pijama}. The M2A system \cite{tang2025towards} was originally trained on the Maestro dataset and subsequently fine-tuned using a curated subset of 371 performances from ATEPP. All three datasets provide performance MIDIs paired with corresponding audio recordings. As presented in Table~\ref{tab:dataset_comparison}, ATEPP and Maestro are classical piano performance corpora, comprising transcribed and recorded performance MIDIs, respectively. The Pijama dataset contains transcribed jazz piano solos recorded in live and studio settings. For evaluation, we use only the test set for ATEPP and randomly select 100 performances from each dataset to ensure diverse compositional styles and recording conditions.

Besides human performance recordings, reconstructed audio from Piano-Encodec is used as an additional reference for calculating metrics. This helps evaluate how well MIDI-VALLE aligns with its training target: the codec representations extracted by Piano-Encodec. All performances are divided into 15-20 second segments, and the metrics are calculated by comparing the model-generated outputs to both the ground truth audio and the Piano-Encodec reconstructions.

% \begin{table}[htb]
% \centering
% \begin{tabular}{c|c|c}
% \hline
% \textbf{Metric} & \textbf{SISDR} & \textbf{VISQOL} \\
% \hline
% \textbf{Encodec} & * & * \\
% \hline
% \textbf{Piano-Encodec} & * & * \\
% \hline
% \end{tabular}
% \caption{Comparison between Encodec and Piano-Encodec models using SISDR and VISQOL metrics.}% \end{table}

\subsection{Listening Tests}
\label{sec:listening_test}
The listening test evaluates the synthesis quality of generations by MIDI-VALLE and M2A, and their compatibility with different EPR systems in a two-stage MPS pipeline, using two types of preference-based evaluations.

For the synthesis quality evaluation, participants are presented with a reference audio recording of human performances alongside two synthesised versions of the same musical excerpt—one generated by MIDI-VALLE and the other by M2A, both conditioned on the same performance MIDI. Participants are asked to select the version that more closely resembles the reference in terms of timbre, phrasing, and expressiveness. The stimuli are drawn from the three datasets used in the objective evaluation, consisting of 6 excerpts from ATEPP, 4 from Maestro, and 4 from Pijama. Each excerpt represents a distinct composition and performance, lasting approximately 15–20 seconds. In total, 14 pairwise comparisons are created for this evaluation.

For the system compatibility evaluation, participants are presented with two synthesised outputs generated by MIDI-VALLE and M2A, based on performance MIDIs produced by different EPR systems. Three EPR systems are considered: M2M \cite{tang2025towards}, a Transformer-based model introduced alongside M2A as part of an MPS system; VirtuosoNet \cite{Jeong2019VirtuosoNetAH}, which employs a hierarchical recurrent neural network (RNN) architecture; and DExter \cite{zhang2024dexter}, a diffusion-based generative model. These systems differ in their MIDI processing and representations, particularly in how sustain pedalling is handled and how MIDI files are encoded. For example, both DExter and VirtuosoNet struggle to accurately model sustain pedal effects, which can result in unnatural note offset predictions. In contrast, M2M tokenises MIDI files but disregards pedalling effects, relying on M2A to synthesise these effects. In our listening test, participants are expected to indicate their preference based on the naturalness, clarity, and expressiveness of the synthesised audio for the same performance MIDI. For each EPR system, four 15–20 second excerpts of distinct compositions are selected, leading to 12 pairwise comparisons.

A total of 20 participants, almost all with over 2 years of music training, were recruited, with each evaluating half of the stimuli. This ensured that each stimulus was assessed by 9 to 11 participants, resulting in a total of 260 votes.

\section{Results \& Discussion}
\label{sec:results}
\subsection{Objective Evaluation}
\label{sec:results_obj}
As shown in Table~\ref{tab:codec_eval}, fine-tuning with the ATEPP dataset significantly enhanced Piano-Encodec compared to the original Encodec \cite{copet2023simple}, reducing spectrogram distortion from 0.304 to 0.123 and chroma distortion from 0.478 to 0.140. In addition, Piano-Encodec achieves high-fidelity reconstruction of human performances, with much lower FAD, spectrogram, and chroma distortions than generative models. Although fine-tuned using the ATEPP dataset, the Piano-Encodec model achieves impressive reconstruction quality on both the Maestro and Pijama datasets. These results validate the reliability of the Piano-Encodec model as an embedding extraction tool for assessing acoustic and musical similarity between synthesised outputs and reference audio. 

\begin{table}[htb]
\centering
\small
\setlength{\tabcolsep}{4pt}
\begin{tabular}{lcccc}
\toprule
\textbf{Model} & \textbf{Dataset} & \textbf{FAD $\downarrow$} & \textbf{Spec. $\downarrow$} & \textbf{Chroma $\downarrow$} \\
\midrule
Encodec \cite{copet2023simple} & ATEPP & -- & 0.304 ± .005 & 0.478 ± .011 \\
\multirow{3}{*}{Piano-Enc.} 
& ATEPP   & 0.685 & 0.123 ± .002 & 0.140 ± .002 \\
& Maestro & 0.984 & 0.135 ± .002 & 0.139 ± .001 \\
& Pijama  & 1.133 & 0.143 ± .003 & 0.137 ± .001 \\
\bottomrule
\end{tabular}
\caption{Reconstruction quality of Encodec \cite{fossez2023high} and Piano-Encodec (Piano-Enc.) evaluated on three datasets. Metrics are calculated by comparing the reconstructed performances with the groundtruth recordings. FAD, spectrogram distance (Spec.), and chroma distance with 95\% confidence intervals are presented.}
\label{tab:codec_eval}
\end{table}

Compared to M2A, as presented in Table~\ref{tab:obj_metrics}, our MIDI-VALLE model achieves over 75\% lower FAD on the ATEPP and Maestro datasets, showing that MIDI-VALLE effectively maps MIDI tokens to realistic audio. However, the high FAD scores for Pijama indicate that MIDI-VALLE struggles with jazz performances, likely due to its training on classical music, which limits its ability to capture the complex harmonic structures and rhythms of jazz. Furthermore, the FAD between MIDI-VALLE outputs and reconstructions is lower than with ground truth, suggesting that MIDI-VALLE aligns more with the quantised representations used in training than with the original audio.

\begin{table}
\centering
\small
\begin{tabular}{lcccc}
\toprule
\textbf{Model} & \textbf{Ref.} & \textbf{FAD $\downarrow$} & \textbf{Spec. $\downarrow$} & \textbf{Chroma $\downarrow$} \\
\midrule
\multicolumn{5}{c}{\textbf{ATEPP}} \\
\cmidrule(lr){1-5}
M2A \cite{tang2025towards} & GT$^1$ & 11.014 & 0.218 ± .005 & 0.421 ± .017 \\
    & RC$^2$ & 11.463 & 0.214 ± .004 & 0.464 ± .017 \\
MV  & GT & 3.329  & 0.219 ± .005 & 0.436 ± .012 \\
    & RC & 2.659  & 0.199 ± .005 & 0.442 ± .012 \\
\midrule
\multicolumn{5}{c}{\textbf{Maestro}} \\
\cmidrule(lr){1-5}
M2A \cite{tang2025towards} & GT & 34.479 & 0.230 ± .003 & 0.387 ± .007 \\
    & RC & 33.753 & 0.224 ± .003 & 0.427 ± .007 \\
MV  & GT & 11.281 & 0.231 ± .004 & 0.428 ± .009 \\
    & RC & 9.168  & 0.206 ± .003 & 0.420 ± .009 \\
\midrule
\multicolumn{5}{c}{\textbf{Pijama}} \\
\cmidrule(lr){1-5}
M2A \cite{tang2025towards} & GT & 274.153 & 0.312 ± .010 & 0.471 ± .009 \\
    & RC & 267.969 & 0.293 ± .008 & 0.509 ± .010 \\
MV  & GT & 102.022 & 0.322 ± .010 & 0.558 ± .014 \\
    & RC & 97.634  & 0.298 ± .009 & 0.584 ± .015 \\
\bottomrule
\end{tabular}
\caption{Spectrogram (Spec.) and chroma distortions are presented along with 95\% confidence intervals for comparing M2A \cite{tang2025towards} and MIDI-VALLE (MV) on the ATEPP, Maestro, and Pijama datasets. Metrics are calculated by comparing the model generations with the reference (Ref.). $^1$GT refers to the groudtruth performance recording and $^2$RC indicates audio reconstructed via Piano-Encodec.}
\label{tab:obj_metrics}
\vspace{-1em}
\end{table}

\begin{figure}[!hbt]
    \centering
\includegraphics[width=.9\linewidth]{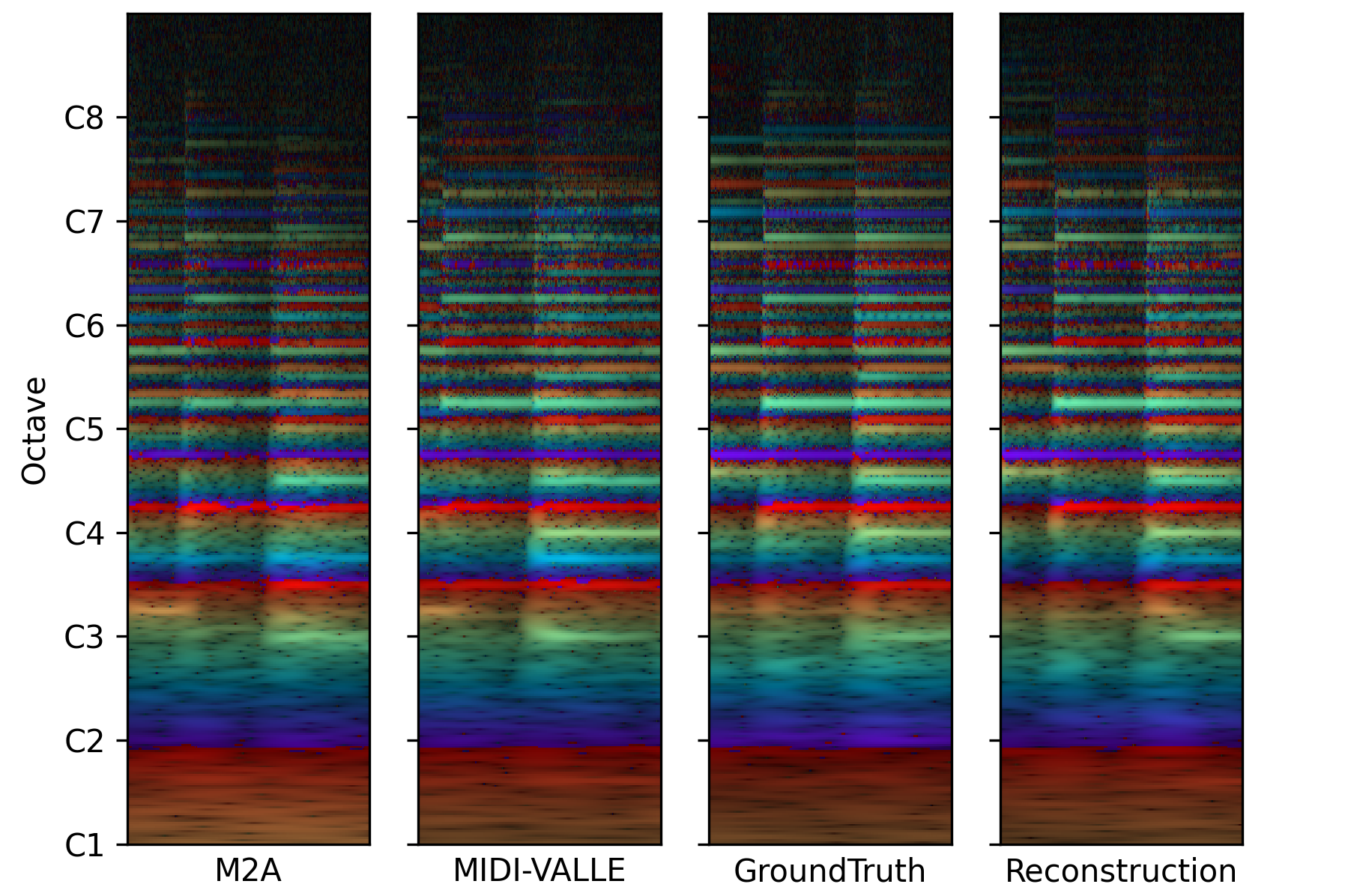}
\caption{Rainbow-grams \cite{shi2023can} of the performances synthesised by MIDI-VALLE and M2A,
along with the ground truth and the Piano-Encodec reconstruction, are shown. The rainbow-gram, based on the Constant-Q Transform, uses colour to represent instantaneous frequency and lightness to indicate spectral amplitude. As the spectral amplitude of a frequency bin increases, the corresponding image pixel becomes lighter.}
\label{fig:rainbow}
\vspace{-1em}
\end{figure}

In terms of chroma distortion, MIDI-VALLE shows similar harmonic consistency to M2A on ATEPP, whereas M2A slightly outperforms MIDI-VALLE on Maestro. This performance difference aligns with the fact that M2A was originally trained on the Maestro dataset. Regarding spectrogram distortion, which reflects the model’s ability to reconstruct acoustics and timbre, MIDI-VALLE exhibits a smaller distance to the reconstruction, compared to M2A on both ATEPP and Maestro. Additionally, as shown in Figure~\ref{fig:rainbow}, MIDI-VALLE provides a more accurate reconstruction across the full frequency spectrum compared to M2A, improving both the timbre realism and the perceptual weight of the sound. These results suggest that MIDI-VALLE effectively adapts to various recording environments and reconstructs acoustics and ambient sound that closely match the provided audio prompts.

Furthermore, MIDI-VALLE, trained solely with transcribed performance MIDIs, generalises well to recorded MIDIs without fine-tuning, as shown by its lower FAD${\text{enc}}$ score and similar spectrogram and chroma distortions to M2A. This makes it beneficial for real-world applications with limited recorded data but rich transcribed data. In contrast, M2A, which is trained on recorded performance MIDIs, struggles to adapt to transcribed datasets without fine-tuning \cite{tang2025towards}, primarily due to the inherent limitations of the piano-roll representation, as discussed in Section~\ref{sec:midi_tokenisation}. 

On the Pijama dataset, MIDI-VALLE exhibits increased spectrogram and chroma distortions, highlighting its difficulty in capturing the complex harmonic structures, syncopated rhythms, and nuanced articulations characteristic of jazz music. These stylistic differences from classical music might lead to unseen token patterns in the MIDI representation, making adaptation challenging. Nevertheless, MIDI-VALLE still outperforms M2A in terms of FAD and achieves comparable spectrogram distortion, suggesting that it better preserves timbral and ambient features that contribute to perceptual similarity. However, the considerable FAD gap between MIDI-VALLE and the ground truth indicates that the overall audio quality remains limited.

\subsection{Subjective Evaluation}
\label{sec:results_sbj}

\begin{figure}[!hbt]
    \centering
\includegraphics[width=.85\linewidth]{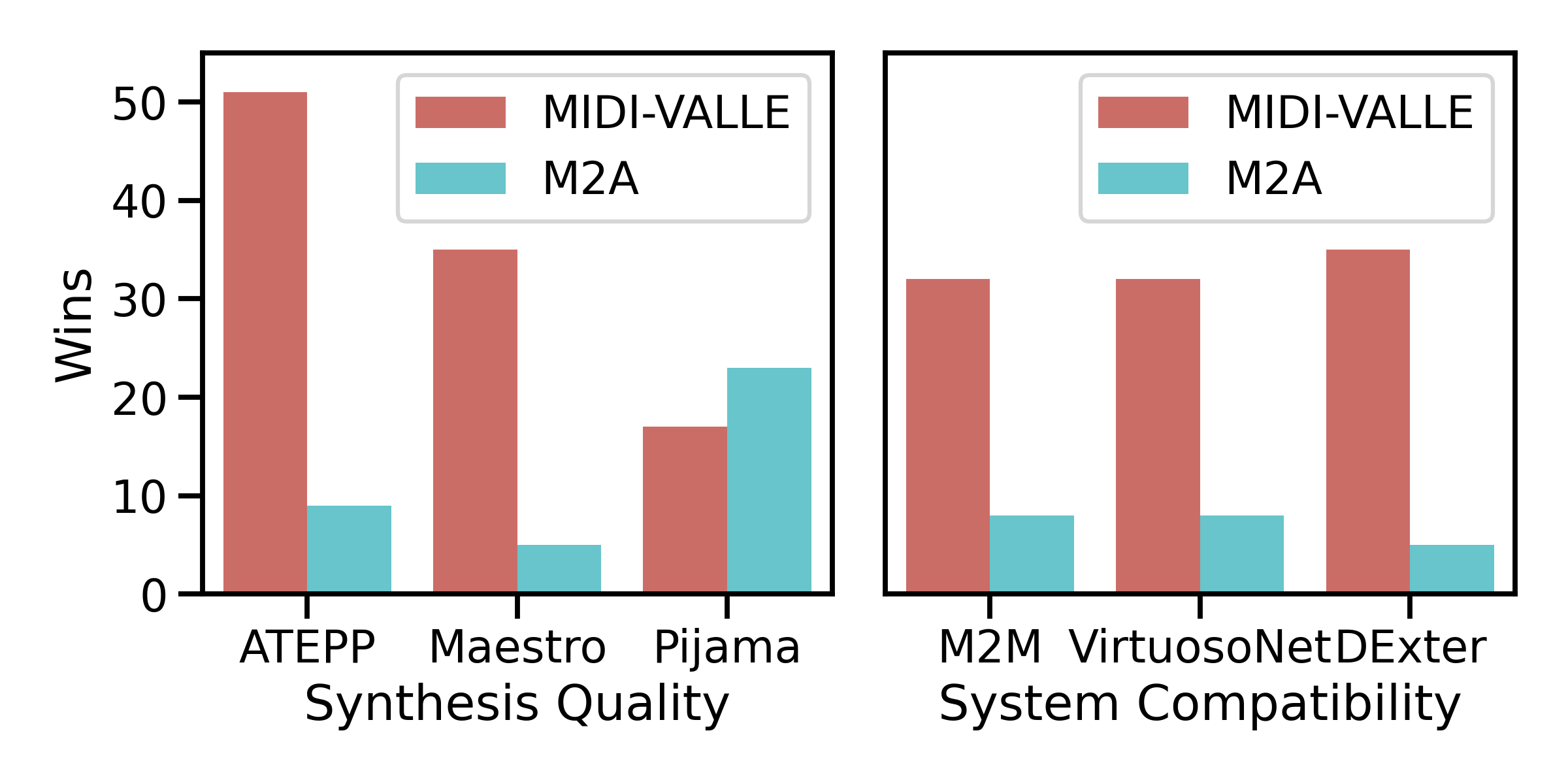}
\caption{Win counts are presented for MIDI-VALLE and M2A models across multiple datasets and combined EPR systems in the listening tests.}
\label{fig:res_lt}
\vspace{-1em}
\end{figure}

The listening test results further validate the findings from the objective metrics. As shown in Figure~\ref{fig:res_lt}, MIDI-VALLE receives significantly more votes than M2A in the synthesis quality evaluation on the ATEPP and Maestro datasets. However, M2A is favoured for segments from Pijama dataset, indicating that while MIDI-VALLE generalises well to classical piano, it requires further refinement to adapt effectively to stylistically distinct genres such as jazz. In the system compatibility evaluation, MIDI-VALLE is consistently preferred over M2A across all EPR systems, demonstrating better adaptability to subtle timing and articulation differences in performance MIDI. While M2A’s piano-roll representation is prone to artefacts under such variations, MIDI-VALLE remains robust, producing more natural and expressive outputs. 

Additionally, the output quality of MIDI-VALLE is strongly influenced by the audio prompt due to its inherited zero-shot design. We observed that, beyond capturing ambient characteristics, the prompt could also determine the loudness and timbre of the generated audio, highlighting the model's ability to adapt to diverse acoustic environments. Moreover, while using the first three seconds of a target segment often produces high-quality results, MIDI-VALLE can generate coherent and natural outputs from any prompt that is stylistically consistent and acoustically clear, enabling it to handle improvised inputs within the classical style.

However, the current design requires precise alignment between MIDI and audio prompts. Subtle timing variations or extra notes can lead to unexpected notes or omissions at the start of the generation. Despite the accurate truncation of the MIDI prompt to three seconds, misalignments still occur. As shown in Figure~\ref{fig:rainbow}, the plot of MIDI-VALLE output appears slightly shifted due to the truncation occurring in the middle of the first note, impacting both its timing and articulation. While manually selecting cutting points that align with the end of MIDI notes, and when the sound completely fades in the audio could resolve these misalignments, this method is not practical for synthesising multiple segments into a complete performance. When generating long performances by concatenating multiple synthesised segments, discontinuities in acoustic characteristics can still be observed.

\section{Conclusion}
We present MIDI-VALLE, a novel EPS model adapted from the VALLE framework, for performance MIDI-to-audio synthesis. Our results demonstrate that MIDI-VALLE outperforms the existing EPS baseline in both adaptability and synthesis quality, producing more natural and expressive audio across a wide range of performance inputs and recording conditions. This improvement is primarily attributed to the discrete tokenisation approach and its inherited zero-shot design, which enhances the model's ability to capture performance nuances and adapt to diverse inputs. Future work will focus on improving generalisation across musical genres, investigating the impact of model size and the role of codebooks in expressive audio generation, and broadening comparisons with physical synthesis methods and alternative audio codec models.
% For BibTeX users:

\section{Acknowledgement}
This work was supported by the UKRI Centre for Doctoral Training in Artificial Intelligence and Music [grant number EP/S022694/1] and the National Institute of Informatics, Japan. J.~Tang is a research student jointly funded by the China Scholarship Council [grant number 202008440382] and Queen Mary University of London. G.~Wiggins received funding from the Flemish Government under the "Onderzoeksprogramma Artificiële Intelligentie (AI) Vlaanderen". We thank the reviewers for their valuable feedback, which helped improve the quality of this work.

\section{Ethics Statement}
No personal or sensitive user data is involved in this research. The datasets used in this study — ATEPP \cite{zhang2022atepp}, Maestro \cite{hawthorne2018enabling}, and Pijama \cite{edwards2023pijama} — contain audio recordings and corresponding MIDI annotations of piano performances. The MIDI files from all three datasets are publicly available. However, the audio recordings in ATEPP and Pijama are accessible exclusively for research purposes under academic use agreements, and have been used accordingly.

All model training and evaluation were conducted in compliance with these terms, with no commercial usage or redistribution of the restricted audio data. The listening tests involved voluntary participation by musically trained individuals, who were informed of the purpose and anonymised participation. No personal data was collected. The study was reviewed and approved by the Electronic Engineering and Computer Science Devolved School Research Ethics Committee at Queen Mary University of London under reference number QMERC20.565.DSEECS25.019.

Code and generated audio examples are made available to promote transparency and reproducibility. We acknowledge the potential for misuse of generative audio models, including the synthesis of deceptive or misleading content. We strongly discourage such applications and advocate for the responsible use of this technology, including clear attribution and disclosure when synthetic audio is employed.
\bibliography{citation}
\end{document}